%% file: VB-Skewness.tex
\documentclass[10pt,twocolumn,twoside] {IEEEtran}
\ifCLASSOPTIONcompsoc \usepackage[caption=false,font=normalsize,labelfon
t=sf,textfont=sf]{subfig}
\else
\usepackage[caption=false,font=footnotesize]{subfig} 
\fi

\input{header.tex}
\newcommand{\figs}{}

\newcommand{\eye}{I}
\newcommand{\ones}{\mathbf{1}}

\title{ Robust Inference for State-Space Models\\ with Skewed Measurement Noise}

\author{Henri Nurminen, Tohid Ardeshiri, Robert Pich\'e,~\IEEEmembership{Member,~IEEE}, and Fredrik Gustafsson,~\IEEEmembership{Fellow,~IEEE}
\thanks{Copyright (c) 2015 IEEE. Personal use of this material is permitted. However, permission to use this material for any other purposes must be obtained from the IEEE by sending a request to pubs-permissions@ieee.org.}
\thanks{H. Nurminen and R. Pich\'e are with the Department of Automation Science and Engineering, Tampere University of Technology (TUT), PO Box 692, 33101 Tampere, Finland  (e-mails: henri.nurminen@tut.fi, robert.piche@tut.fi). H. Nurminen receives funding from TUT Graduate School, Finnish Doctoral Programme in Computational Sciences (FICS), and the Foundation of Nokia Corporation.}
\thanks{T. Ardeshiri and F. Gustafsson are with the Department of Electrical Engineering, Link\"{o}ping University, 58183 Link\"{o}ping, Sweden, (e-mails: tohid@isy.liu.se, fredrik@isy.liu.se). T. Ardeshiri receives funding from  Swedish research council (VR), project ETT (621-2010-4301)}
}

\begin{document}
\maketitle

\begin{abstract}
Filtering and smoothing algorithms for linear discrete-time state-space models with skewed and heavy-tailed measurement noise are presented. The algorithms use a variational Bayes approximation of the posterior distribution of models that have normal prior and skew-$t$-distributed measurement noise. The proposed filter and smoother are compared with conventional low-complexity alternatives in a simulated pseudorange positioning scenario. In the simulations the proposed methods achieve better accuracy than the alternative methods, the computational complexity of the filter being roughly 5 to 10 times that of the Kalman filter.
\end{abstract}


\input{introduction}
\input{formulation}
\input{solution}
\input{simulations}
\input{conclusions}
\vfill\clearpage
\bibliographystyle{IEEEtran}
\bibliography{IEEEabrv,VB-Skewness}

\end{document}

%% file: header.tex
\usepackage{amsmath,graphicx}
\usepackage[english]{babel}
\usepackage[latin1]{inputenc}
\usepackage{tikz}

\newcommand{\displaycomments}

\ifdefined\displaycomments
\newcommand{\note}[1]{\textcolor{red}{\emph{#1}}}
\else
\newcommand{\note}[1]{}
\fi

\usepackage{xspace}
\usepackage{amsmath}
\usepackage{amsthm}
\usepackage{amssymb}
\usepackage{graphicx}      
\usepackage{algorithmicx} 
\usepackage{cite} 

\usetikzlibrary{arrows}
\usepackage{pgfplots}

\DeclareMathOperator*{\argmin}{argmin}

\newcommand{\E}{\mathop{\mathbb E}}

\renewcommand{\t}[1]{\mathrm{T}#1}
\newcommand{\N}{\mathcal{N}}

\renewcommand{\baselinestretch}{0.9} 
\usepackage{cite} 
\usepackage{stfloats} 
\usepackage{amsmath}

\renewcommand{\d}[1]{\;\mathrm{d}#1}
\renewcommand{\t}[1]{\mathrm{T}#1}

\newcommand{\ST}{\operatorname{ST}}
\newcommand{\T}{\operatorname{T}}
\newcommand{\St}{\operatorname{t}}

\newcommand{\G}{\mathcal{G}}

\renewcommand{\d}[1]{\;\mathrm{d}#1}
\renewcommand{\t}[1]{\mathrm{T}#1}

\usepackage{booktabs}
\usepackage {ftnxtra}
\usepackage{algpseudocode}
\usepackage{epstopdf}

%% file: introduction.tex

\section{Introduction} \label{sec:introduction}

The Kalman filter (KF) \cite{Kalman60} is the linear minimum mean-square-error filter for linear state-space models, but it is optimal within the set of all filters only when the noise processes are normally distributed \cite{anderson1979}. However, the normal distribution has small tail probabilities, and real-world data typically contain large errors (``outliers'') more often than the normal distribution predicts \cite{pearson2002}. Therefore, the KF is prone to large estimation errors when outliers occur. Hence, there is a need for filtering and smoothing algorithms that mitigate the outlier measurements' influence.

\begin{figure}[t]
\centering
\includegraphics[trim=0mm 0mm 0mm 0mm, clip, width=1\columnwidth]{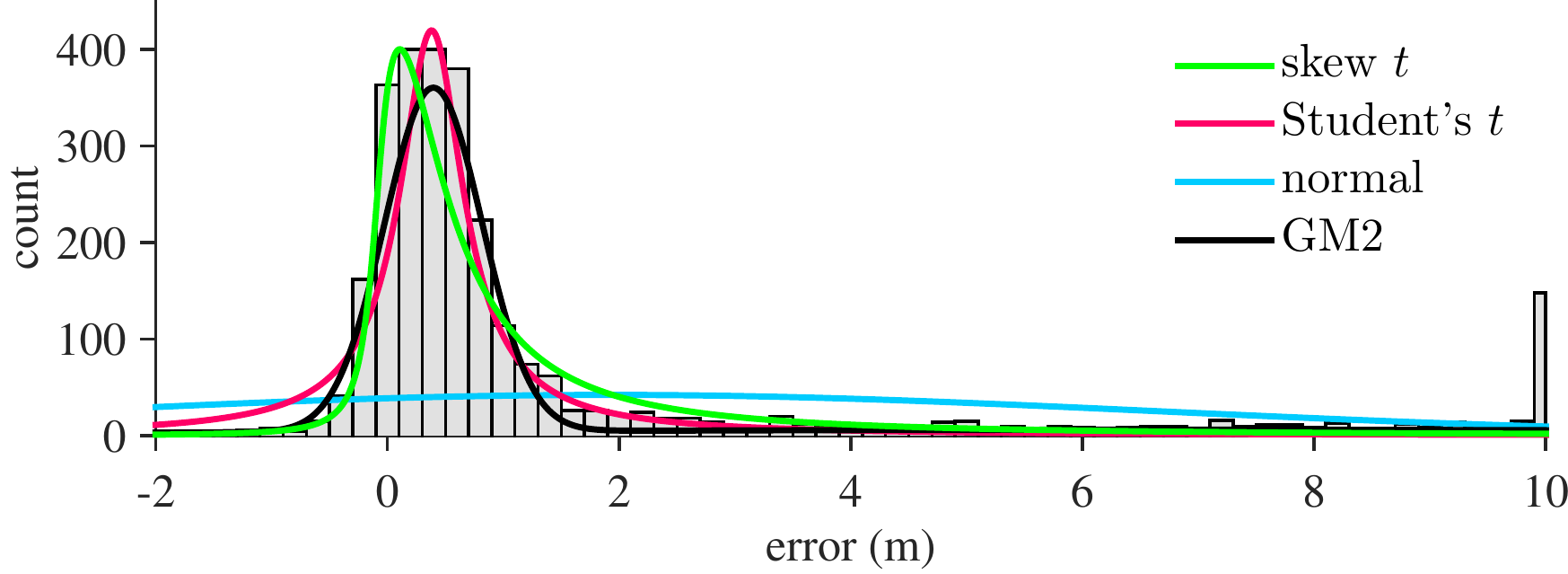}
\vspace{-0.5mm}
\caption{Skewed distributions fit better than symmetric distributions to the time-of-flight measurement errors. BIC values for 2905 data points are 9600 for skew $t$, 10500 for Student's $t$, 17200 for normal, and 10000 for GM2.}
\label{fig:uwb_example}
\end{figure}

Many applications involve noise processes that have both heavy-tailed (high-kurtosis) and asymmetric (skewed) distributions. In radio signal based distance estimation \cite{GusGun2005,BorsenChen2009}, for example, 
non-line-of-sight causes large positive errors  
\cite{kaemarungsi2012,Kok2015}. Fig.\ \ref{fig:uwb_example} shows the error histogram of a time-of-flight based ultra-wideband distance measurement experiment\footnote{High accuracy reference measurements are provided through the use of the
Vicon real-time tracking system courtesy of the UAS Technologies Lab,
Artificial Intelligence and Integrated Computer Systems Division (AIICS)
at the Department of Computer and Information Science (IDA).
http://www.ida.liu.se/divisions/aiics/aiicssite/index.en.shtml
} and maximum likelihood fits of some probability distribution families. 
By the Bayesian information criterion (BIC) \cite{schwarz1978}, the skewed distributions skew $t$ \cite[Ch. 4.3]{azzalini2014} and two-component Gaussian mixture (GM2) model the data better than the symmetric Student's $t$ \cite[Ch. 28]{johnson1995} and normal. Other applications for asymmetric distributions have emerged in biostatistics~\cite{fruhwirth2010}, psychiatry~\cite{eling2012}, environmetrics~\cite{counsell2010}, and economics~\cite{marchenko2010phd}. 

Despite these applications, a computationally efficient estimation algorithm for time-series data with heavy-tailed and asymmetric noise has been missing. Robust algorithms that model the heavy-tailed noise with a $t$-distribution are proposed in \cite{Agamennoni12, piche2012, roth2013}, but these do not use the skewness information. A GM2 can model skewness, but the number of mixture components in the posterior increases exponentially with the number of measurements. Furthermore, the GM2 has heavy tails only within a limited range near the component locations, and it has five parameters, while four suffices for modeling location, spread, skewness and kurtosis. Particle filters (PF) \cite{doucet2000} can cope with a wide range of models including skewed noise processes, but their computational complexity increases rapidly as the state dimension increases.

This letter proposes approximations to the Bayesian filter and smoother that retain the computational efficiency of the KF while introducing more modeling flexibility for skewed and heavy-tailed measurement noise. The measurement noise is modelled by the skew $t$-distribution, and the proposed algorithms use a variational Bayes (VB) approximation of the posterior. The proposed filter and smoother are evaluated by numerical pseudorange positioning simulations, where they are compared with the state-of-the-art computationally light algorithms and a PF. To our knowledge, the only earlier work applying VB approximations to the skew $t$-distribution is that of Wand et al.\ \cite{wand2011}. However, Wand et al.\ do not consider state-space models and time-series estimation.

\section{Skew $t$-distribution}
\begin{figure}[t]
\centering
\includegraphics[trim=0mm 0mm 0mm 0mm, clip, width=\columnwidth]{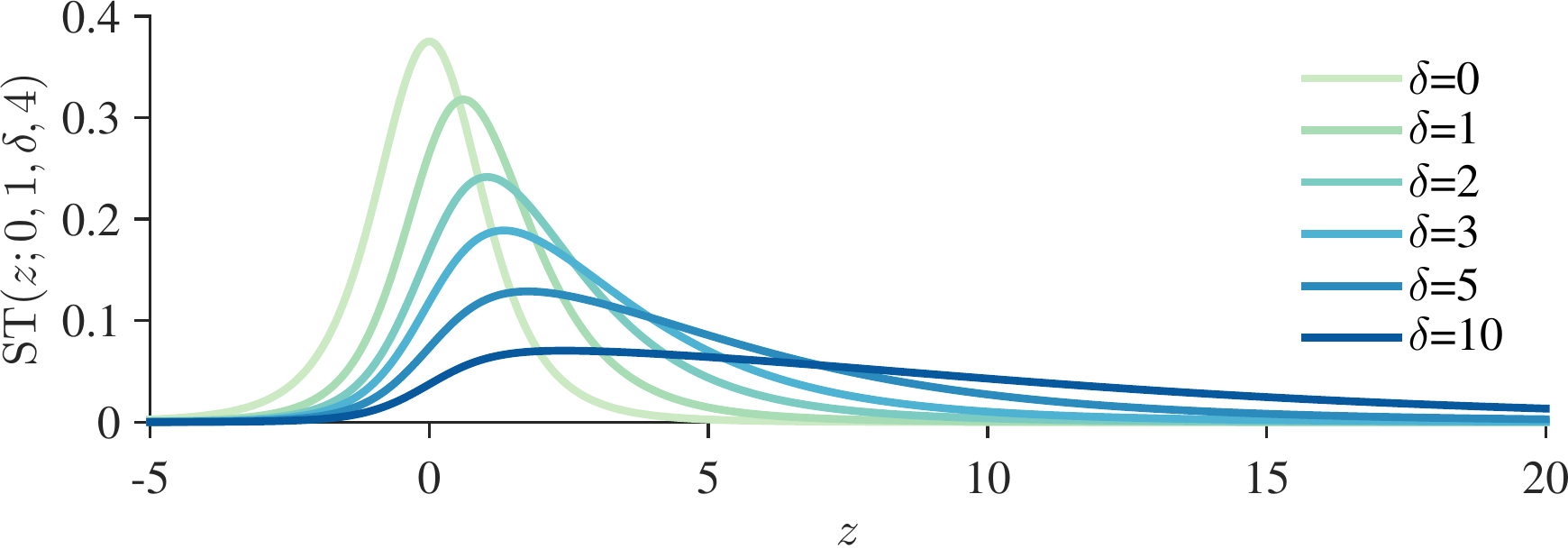}
\caption{The PDF $\ST(z; 0, 1,\delta,4)$ for different shape parameter values $\delta$.} \label{fig:skew-t-delta}
\end{figure}

Skewed extensions of the well-known unimodal symmetric distributions have been studied since the introduction of the skew normal distribution by Azzalini in \cite{azzalini1985}. 
The univariate skew $t$-distribution is parametrized by its location parameter $\mu\in\mathbb{R}$, spread parameter $\sigma>0$, shape parameter $\delta \in\mathbb{R}$  and degrees of freedom $\nu>0$, and has a probability density function (PDF) of the form
\begin{align}
\ST(z; \mu, \sigma^2,\delta,\nu)=2 \St(z;\mu,\delta^2+\sigma^2,\nu)\T(\widetilde{z};0,1,\nu+1)\label{eq:skewt},
\end{align}
where 
\begin{align}
\St(z;\mu,\sigma^2,\nu)=\frac{\Gamma\left(\frac{\nu+1}{2}\right)}{\sigma\sqrt{\nu\pi}\Gamma\left(\frac{\nu}{2}\right)}\left(1+\frac{(z-\mu)^2}{\nu \sigma^2}\right)^{-\frac{\nu+1}{2}}
\end{align}
is the PDF of Student's $t$-distribution, $\Gamma(\cdot)$ is the gamma function, and 
$\widetilde{z} = \frac{(z-\mu)\delta}{\sigma} \left({\frac{\nu+1}{\nu(\delta^2+\sigma^2)+(z-\mu)^2}}\right)^{\frac{1}{2}}$.
Also, $\T(\cdot;0,1,\nu)$ denotes the cumulative distribution function (CDF) of Student's $t$-distribution with degrees of freedom $\nu$.
The PDF $\ST(z; 0, 1,\delta,4)$  is plotted for six different values of shape parameter $\delta$ in Fig. \ref{fig:skew-t-delta}. The skew $t$-distribution approaches normal distribution when $\nu \rightarrow \infty$ and $\delta \rightarrow 0$.
Expressions for the first two moments of the univariate skew $t$-distribution with the parametrisation \eqref{eq:skewt} can be found in \cite{sahu2009}.

Following the introduction of the multivariate skew normal distribution in \cite{azzalini1996}, multivariate skew $t$-distributions have been proposed in \cite{branco2001,azzalini2003, gupta2003skew}. In these versions, the PDF of the skew $t$-distribution involves only the univariate CDF of $t$-distribution, while the definition of skew $t$-distribution given in \cite{sahu2003,lin2010,lee2013} involves the multivariate CDF, but a single kurtosis factor. In this letter the measurement noise distribution is a product of independent univariate skew $t$-distributions. This less general model is justified in applications where one-dimensional data from different sensors can be assumed to be statistically independent.

%% file: formulation.tex

\section{Problem formulation} \label{sec:problem_formulation}
Consider the linear state-space model with skew-$t$-distributed measurement noise
\begin{subequations}
\label{eq:statemodel}
\begin{align}
x_{k+1}&=\ Ax_k+w_k, & w_k&\stackrel{\text{iid}}{\sim}\N(w_k;0,Q),\\
y_{k}&= C x_k+e_k, & [e_k]_i&\stackrel{\text{iid}}{\sim} \ST([e_k]_i;0,R_{ii},\Delta_{ii},\nu_i) \label{eq:measurementmodel}
\end{align}
\end{subequations}
where 
 $\N(\cdot;\mu,\Sigma)$ denotes a (multivariate) normal PDF with mean $\mu$ and covariance $\Sigma$;
 $A \in \mathbb{R}^{n_x\times n_x}$ is the state transition matrix;
 $x_k\in\mathbb{R}^{n_x}$ indexed by $1\leq k\leq K$ is the state to be estimated with prior distribution 
\begin{align}
p(x_1)=\N(x_1;x_{1|0},P_{1|0})\label{eq:prior};
\end{align}
where the subscript ``$a|b$" is read ``at time $a$ using measurements up to time $b$";
 $y_k\in\mathbb{R}^{n_y}$ also indexed by $1\leq k\leq K$ are the measurements and the elements of $y_k$ are conditionally independently skew-$t$-distributed; 
 $R\in\mathbb{R}^{n_y\times n_y}$ is a diagonal matrix whose diagonal elements $R_{ii}$ are the squares of the spread parameters of \eqref{eq:measurementmodel};
 $\Delta\in\mathbb{R}^{n_y\times n_y}$ is a diagonal matrix whose diagonal elements $\Delta_{ii}$ are the shape parameters of \eqref{eq:measurementmodel};
 $\nu\in\mathbb{R}^{n_y}$ is a vector whose elements $\nu_i$ are the degrees of freedom of \eqref{eq:measurementmodel};
 $C \in \mathbb{R}^{n_y\times n_x}$ is the measurement matrix;
 $\{w_k\in\mathbb{R}^{n_x}| 1\leq k \leq K\}$  and $\{e_k\in\mathbb{R}^{n_y}| 1\leq k \leq K\}$  are mutually independent noise sequences;  and the operator $[\cdot]_{ij}$ gives the $(i,j)$ entry of its argument. 

The aim of this letter is to derive a Bayesian filter and a Bayesian smoother using the VB method that computes an approximation of the filtering distribution $p(x_k|y_{1:k})$ and smoothing distribution $p(x_k|y_{1:K})$.

%% file: solution.tex
\section{Variational solution} \label{sec:variational_solution}

The likelihood function implied from~\eqref{eq:measurementmodel} has the hierarchical representation \cite{lin2010}
\begin{subequations}
\label{eq:hierarchical}
\begin{align}
y_k|x_k,u_k,\Lambda_k &\thicksim \N(Cx_k+\Delta u_k,\Lambda_k^{-1}R),\\
u_k|\Lambda_k &\thicksim \N_+(0,\Lambda_k^{-1}),\\
[\Lambda_k]_{ii} &\thicksim \G\left(\frac{\nu_i}{2},\frac{\nu_i}{2}\right) .
\end{align}
\end{subequations}
$\Lambda_k$ is a diagonal matrix with independent random diagonal elements $[\Lambda_k]_{ii}$, and $\N_+(\mu,\Sigma)$ denotes the (multivariate) truncated normal distribution with closed positive orthant as support, location parameter $\mu$, and squared-scale matrix $\Sigma$. Furthermore, $\G(\alpha, \beta)$ denotes the gamma distribution with shape parameter $\alpha$ and rate parameter $\beta$.

Using Bayes' theorem, the likelihood \eqref{eq:hierarchical} and the prior~\eqref{eq:prior},  the joint smoothing posterior PDF can be written as
\begin{align}
p(x_{1:K}&,u_{1:K},\Lambda_{1:K}|y_{1:K})\propto p(x_1)\prod_{l=1}^{K-1} p(x_{l+1}|x_l)\nonumber\\
&\hspace{1cm}\times\prod_{k=1}^K p(y_k|x_k,u_k,\Lambda_k)p(u_k|\Lambda_k)p(\Lambda_k) \\
=&\N(x_1;x_{1|0},P_{1|0})\prod_{l=1}^{K-1}\N(x_{l+1};Ax_l,Q) \nonumber\\
&\hspace{2mm}\times\prod_{k=1}^K  \N(y_k;Cx_k+\Delta u_k,\Lambda_k^{-1}R) \N_+(u_k; 0,\Lambda_k^{-1})\nonumber\\
&\hspace{2mm}\times\prod_{k=1}^K\prod_{i=1}^{n_y}\G\left([\Lambda_k]_{ii};\frac{\nu_i}{2},\frac{\nu_i}{2}\right).
\end{align}
This posterior is not analytically tractable. We seek an approximation in the form
\begin{align}
\label{eq:factors}
p(x_{1:K},&u_{1:K},\Lambda_{1:K}|y_{1:K})\approx q_x(x_{1:K})q_u(u_{1:K})q_\Lambda(\Lambda_{1:K}) .
\end{align}
In the VB approach, the Kullback-Leibler divergence (KLD)~\cite{CoverT2006}  of the true posterior from the factorized approximation is minimized;
\begin{align}
&\hat{q}_{x},\hat{q}_{u},\hat{q}_{\Lambda}=\argmin_{{q}_{x},{q}_{u},{q}_{\Lambda}}\nonumber\\
&D_{\text{KL}}(q_x(x_{1:K})q_u(u_{1:K})q_\Lambda(\Lambda_{1:K})||p(x_{1:K},u_{1:K},\Lambda_{1:K}|y_{1:K}))\nonumber
\end{align}
where $D_{\text{KL}}(q(\cdot)||p(\cdot))\triangleq\int q(x)\log\frac{q(x)}{p(x)}\d x$ is the KLD.  
The analytical solutions for $\hat{q}_x$, $\hat{q}_u$ and $\hat{q}_\Lambda$ can be obtained by cyclic iteration of
\small{
\begin{subequations}
\label{eqn:IterativeOptimization}
\begin{align}
\log {q}_{x}(x_{1:K}) \leftarrow& \E_{{q}_{u}{q}_{\Lambda}}[\log p(y_{1:K},x_{1:K},u_{1:K},\Lambda_{1:K})]+c_{x}\label{eqn:IterativeOptimizationx}\\
\log {q}_{u}(u_{1:K}) \leftarrow& \E_{{q}_{x}{q}_{\Lambda}}[\log p(y_{1:K},x_{1:K},u_{1:K},\Lambda_{1:K})]+c_{u}\label{eqn:IterativeOptimizationu}\\
\log {q}_{\Lambda}(\Lambda_{1:K}) \leftarrow& \E_{{q}_{x}{q}_{u}}[\log p(y_{1:K},x_{1:K},u_{1:K},\Lambda_{1:K})]+c_{\Lambda}\label{eqn:IterativeOptimizationL}
\end{align}
\end{subequations}}\normalsize
where the expected values on the right hand sides of~\eqref{eqn:IterativeOptimization} are taken with respect to the current $q_x$, $q_u$ and $q_\Lambda$~\cite[Chapter 10]{Bishop2007}\cite{TzikasLG2008,Beal03}. Also, $c_{x}$, $c_u$ and $c_\Lambda$  are constants with respect to the variables $x_k$, $u_k$ and $\Lambda_k$,  respectively. This recursion is convergent to a local optimum~\cite[Chapter 10]{Bishop2007}. When the iterations converge, approximate densities $q_u$ and $q_\Lambda$ are integrated out from the right hand side of~\eqref{eq:factors}  by simply discarding them. Then, the approximate marginal smoothing density $q_x(x_k)$ is obtained, and it turns out to be a normal distribution 
$q_x(x_k)=\N(x_k;x_{k|K},P_{k|K})$ 
where the parameters $x_{k|K}$ and $P_{k|K}$ are the output of the smoothing algorithm given in Table~\ref{table:smoothing}. The filtering algorithm and the parameters of the filtering posterior $q_x(x_k)=\N(x_k;x_{k|k},P_{k|k})$ can be found in Table~\ref{table:filtering}. The  derivations for the expectations given in \eqref{eqn:IterativeOptimization} are relegated to \cite{Ardeshiri15techreport} because of space constraints. 

\begin{table}
\caption{Smoothing for skew-$t$ measurement noise}\label{table:smoothing}
\vspace{-5mm}\rule{\columnwidth}{1pt}
\begin{algorithmic}[1]
\State \textbf{Inputs:} $A$, $C$, $Q$, $R$, $\Delta$,  $\nu$, $x_{1|0}$,  $P_{1|0}$ and $y_{1:K}$
\Statex \hspace{0mm}\textit{initialization}
\State $\overline{\Lambda_{k}} \gets I_{n_y}$ for $k=1\cdots K$
\State $\overline{u_{k}}\gets 0$ for $k=1\cdots K$
\Repeat
\Statex \hspace{2mm}\textit{update $q_x(x_{1:K})$ given $q_u(u_{1:K})$ and $q_\Lambda(\Lambda_{1:K})$}
	\For{$k$ = 1 to K }
		\State $K_x\gets P_{k|k-1}C^\t(CP_{k|k-1}C^\t+\overline{\Lambda_k}^{-1}R)^{-1}$
		\State $x_{k|k}\gets x_{k|k-1}+K_x(y_k-Cx_{k|k-1}-\Delta \overline{u_k})  $
		\State $P_{k|k}\gets (I-K_xC)P_{k|k-1}$
    \Statex \hspace{6mm}\textit{predict $q_x(x_{k+1})$}
	  \State $x_{k+1|k} \gets Ax_{k|k}$
	  \State $P_{k+1|k} \gets AP_{k|k}A^\t+Q$	
	\EndFor
	\For{$k$ = K-1 down to 1 }
		\State $G_k\gets P_{k|k}A^\t P_{k+1|k}^{-1}$
		\State $x_{k|K}\gets x_{k|k}+G_k(x_{k+1|K}-Ax_{k|k})$
		\State $P_{k|K}\gets P_{k|k}+G_k(P_{k+1|K}-P_{k+1|k})G_k^\t$
	\EndFor
	\Statex \hspace{4mm}\textit{update $q_u(u_{1:K})$  and $q_\Lambda(\Lambda_{1:K})$ given  $q_x(x_{1:K})$}
	\For{$k$ = 1 to K }
		\Statex \hspace{6mm}\textit{update $q_u(u_k)=\N_+(u_k;u_{k|K},U_{k|K})$ }
		\State $\widetilde{u}_k=y_k-Cx_{k|K}$
		\State $K_u \gets \Delta(\Delta^2+ R)^{-1}$
		\State $u_{k|K}\gets K_u \widetilde{u}_k$
		\State $U_{k|K}\gets (I-K_u\Delta)\overline{\Lambda_k}^{-1}$
		\State $\overline{u_k}\gets \E_{\N_+(u_{k|K},U_{k|K})}[u_k]$ \Comment{ see  \cite{Barr1999} for the formula}
		\For{$i$ = 1 to $n_y$ }
			\State $\Upsilon_{ii}\gets\E_{\N_+(u_{k|K},U_{k|K})}[ [u_k]_i^2] $ \Comment{ see \cite{Barr1999} for the formula}
		\EndFor		\Statex \hspace{6mm}\textit{update $q_\Lambda(\Lambda_k)=\prod_{i=1}^{n_y}\G\left([\Lambda_k]_{ii};\frac{\nu_i}{2}+1,\frac{\nu_i+[\Psi_k]_{ii}}{2}\right)$ }
		\State  $\Psi_k\gets    R^{-1}(\widetilde{u}_k \widetilde{u}_k^\t+CP_{k|K}C^\t)+  (\Delta R^{-1}\Delta+I)\Upsilon$
		\Statex $\hspace{1.5cm}    -R^{-1}\Delta \overline{u_k}\widetilde{u}_k^\t - \Delta R^{-1}\widetilde{u}_k \overline{u_k}^\t$
		\State $[\overline{\Lambda_k}]_{ii} \gets\frac{\nu_i+2}{\nu_i+[\Psi_k]_{ii}}$
	\EndFor	
\Until{\textbf{converged}}
\State \textbf{Outputs: $x_{k|K}$ and  $P_{k|K}$ for $k=1\cdots K$ } 
\end{algorithmic}
\noindent \rule{\columnwidth}{1pt}\vspace{0mm}
\end{table}
\begin{table}
\caption{Filtering for skew-$t$ measurement noise}\label{table:filtering}
\vspace{-5mm}\rule{\columnwidth}{1pt}
\begin{algorithmic}[1]
\State \textbf{Inputs:} $A$, $C$, $Q$, $R$, $\Delta$,  $\nu$, $x_{1|0}$,  $P_{1|0}$ and $y_{1:K}$
\For{$k$ = 1 to K }
	\Statex \hspace{2mm}\textit{initialization}
	\State $\overline{\Lambda_k} \gets I_{n_y}$
	\State $\overline{u_k}\gets 0$
	\Repeat
		\Statex \hspace{6mm}\textit{update $q_x(x_k)=\N(x_k;x_{k|k},P_{k|k})$ given $q_u(u_k)$ and $q_\Lambda(\Lambda_k)$}
		\State $K_x\gets P_{k|k-1}C^\t(CP_{k|k-1}C^\t+\overline{\Lambda_k}^{-1}R)^{-1}$
		\State $x_{k|k}\gets x_{k|k-1}+K_x(y_k-Cx_{k|k-1}-\Delta \overline{u_k})  $
		\State $P_{k|k}\gets (I-K_xC)P_{k|k-1}$
		\Statex \hspace{6mm}\textit{update $q_u(u_k)=\N_+(u_k;u_{k|k},U_{k|k})$ given $q_x(x_k)$ and $q_\Lambda(\Lambda_k)$}
		\State $K_u \gets \Delta(\Delta^2+ R)^{-1}$
		\State $\widetilde{u}_k=y_k-Cx_{k|k}$
		\State $u_{k|k}\gets K_u \widetilde{u}_k$
		\State $U_{k|k}\gets (I-K_u\Delta)\overline{\Lambda_k}^{-1}$
		\State $\overline{u_k}\gets \E_{\N_+(u_{k|k},U_{k|k})}[u_k]$ \Comment{ see \cite{Barr1999} for the formula}
		\For{$i$ = 1 to $n_y$ }
			\State $\Upsilon_{ii}\gets\E_{\N_+(u_{k|k},U_{k|k})}[ [u_k]_i^2] $ \Comment{ see \cite{Barr1999} for the formula}
		\EndFor
		\Statex \hspace{6mm}\textit{update $q_\Lambda(\Lambda_k)=\prod_{i=1}^{n_y}\G\left([\Lambda_k]_{ii};\frac{\nu_i}{2}+1,\frac{\nu_i+[\Psi_k]_{ii}}{2}\right)$ }
		\Statex \hspace{6mm}\textit{ given $q_u(u_k)$ and $q_x(x_k)$}
		\State  $\Psi_k\gets    R^{-1}(\widetilde{u}_k \widetilde{u}_k^\t+CP_{k|k}C^\t)+  (\Delta R^{-1}\Delta+I)\Upsilon $
		\Statex $\hspace{1.5cm}    -R^{-1}\Delta \overline{u_k}\widetilde{u}_k^\t - \Delta R^{-1}\widetilde{u}_k \overline{u_k}^\t$
		\State $[\overline{\Lambda_k}]_{ii} \gets\frac{\nu_i+2}{\nu_i+[\Psi_k]_{ii}}$
		\Until{\textbf{converged}}
  \Statex \hspace{2mm}\textit{predict $q_x(x_{k+1})$}
	\State $x_{k+1|k} \gets Ax_{k|k}$
	\State $P_{k+1|k} \gets AP_{k|k}A^\t+Q$	
\EndFor
\State \textbf{Outputs: $x_{k|k}$ and  $P_{k|k}$ for $k=1\cdots K$ } 
\end{algorithmic}
\noindent \rule{\columnwidth}{1pt}\vspace{0mm}
\end{table}

%% file: simulations.tex

\section{Simulations} \label{sec:simulations}

Numerical simulations are carried out to evaluate the performance of the proposed algorithms Skew-$t$ variational Bayes filter (STVBF) and Skew-$t$ variational Bayes smoother (STVBS). The compared filters are $t$ variational Bayes filter (TVBF) \cite{piche2012}, the bootstrap Particle filter (PF), the Kalman filter (KF), and the KF with measurement validation gating (KF-G) \cite[Ch. 5.7.2]{bar-shalom} that discards the individual measurement components whose normalized squared innovation is larger than the $\chi_1^2$-distribution's 99\,\% quantile. The smoothers are $t$ variational Bayes smoother (TVBS) \cite{piche2012}, and Rauch-Tung-Striebel smoother with gating (RTSS-G)\cite{RTS-1965}.
KF and RTSS use the true mean and covariance of the measurement noise distribution, and the TVBF and TVBS use the true mean and $(\nu-2)/\nu$ times the true covariance as the shape matrix. The computations are done using \textsc{Matlab}.


\subsection{One-dimensional positioning}

The simulation consists of 1000 100-step random-walks of model \eqref{eq:statemodel} with parameters $A=1$, $Q=1$, $C=\ones_{3\times1}$, $R=\eye_{3\times3}$, $\nu=4\cdot\ones_{3\times1}$, and $\Delta=5\cdot\eye_{3\times3}$, where $\ones$ is a vector of ones. The VB iterations of STVBF and TVBF are terminated when the change in the estimate is less than 0.01.

\begin{figure}[ht]
\centering
\includegraphics[width=\columnwidth]{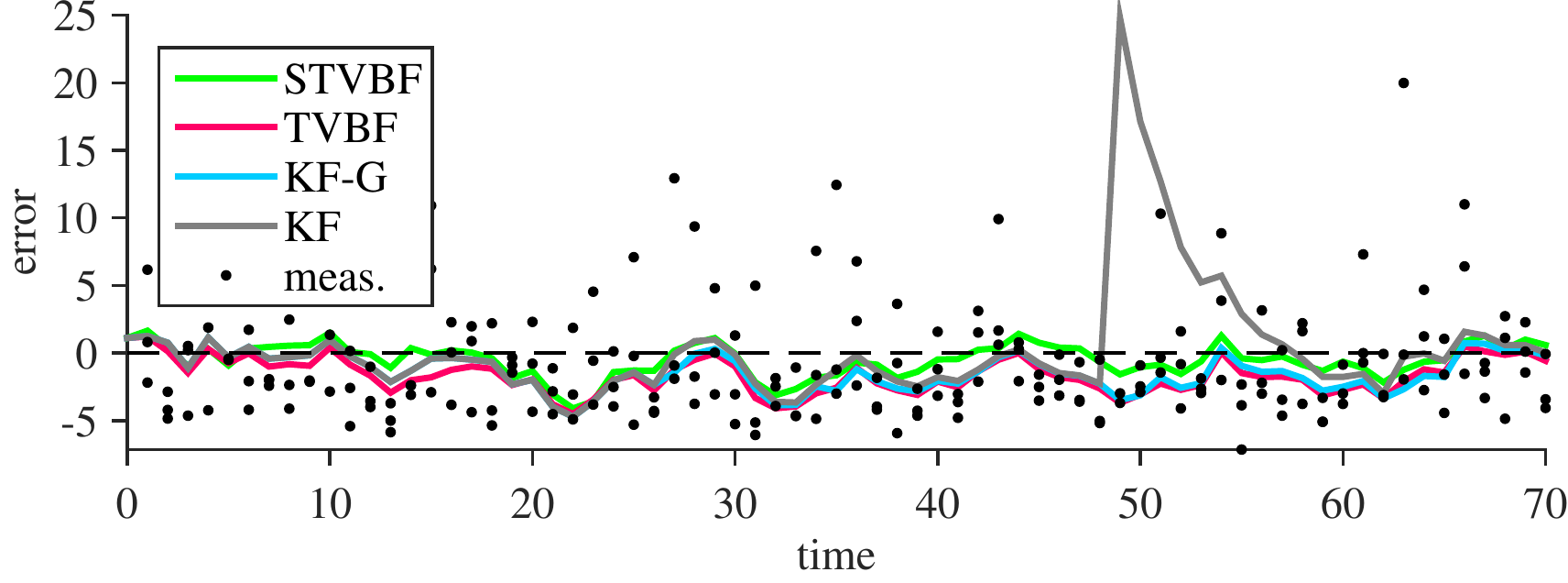}
\vspace{-6mm}
\caption{One-dimensional positioning example illustrates TVBF estimate's negative bias and KF's sensitivity to outliers. Measurement error of 300 at time instant 49 is not shown.}
\label{fig:estimation_example}
\end{figure}

Some statistics of the estimation error are in Table \ref{tab:errmoments}, and Fig.\ \ref{fig:estimation_example} shows an example of the error processes. Table \ref{tab:errmoments} shows that the STVBF has the lowest root-mean-square error (RMSE), the TVBF and KF-G have negative bias, and the KF's error process has the highest standard deviation and positive skew. As illustrated by Fig.\ \ref{fig:estimation_example}, the TVBF and KF-G react relatively slowly to positive errors, interpreting them as outliers to be discounted. The KF error's skewness is caused by excessive sensitivity to the large positive measurement errors.
\begin{table}[ht]
\centering
\caption{Error statistics in one-dimensional positioning}
\label{tab:errmoments}
\begin{tabular}{c|cccc}
Filter & RMSE & Mean & Standard deviation & Skewness \\
\hline
\rule{0pt}{2ex} 
STVBF & 1.2 & 0.1 & 1.2 & 0.0 \\
TVBF & 1.5 & -0.8 & 1.3 & 0.2 \\
KF-G & 1.5 & -0.5 & 1.4 & 0.1 \\
KF & 1.6 & 0.0 & 1.6 & 0.5 
\end{tabular}
\end{table}

\subsection{Pseudorange positioning}

GNSS-type (global navigation satellite system) pseudorange measurements are simulated from the model
\begin{equation} \label{eq:measmodel}
[y_k]_i = \left\| s_i - [x_k]_{1:3} \right\| + [x_k]_4 + [e_k]_i ,\ [e_k]_i \stackrel{\text{iid}}{\thicksim} \mathrm{ST}(0, 1, \delta, 4)
\end{equation}
where $s_i$ is the $i$th satellite's position, $[x_k]_4$ is bias, $e_k $ is noise, and $\delta$ is varied. The model is linearized, and the linearization error is negligible because the satellites are far from the receiver. The state model is a three-dimensional random walk with process noise covariance matrix $Q = \mathrm{diag}( q^2, q^2, 0.5^2 )$, where $q$ is a parameter. The constant bias $[x_k]_4$ has prior $\N(0,0.75^2)$.
Satellite constellations of Global Positioning System provided by the International GNSS service \cite{dow2009} are used, and on average 7.6 satellites are measured.
The results are based on 1000 Monte Carlo replications of a 100-step trajectory. The RMSE is computed for the components $[x_k]_{1:3}$.

\subsubsection{Evaluation of the filter}
\begin{figure}[t]
\centering
\subfloat[$\delta=2$]{\includegraphics[width=0.48 \columnwidth]{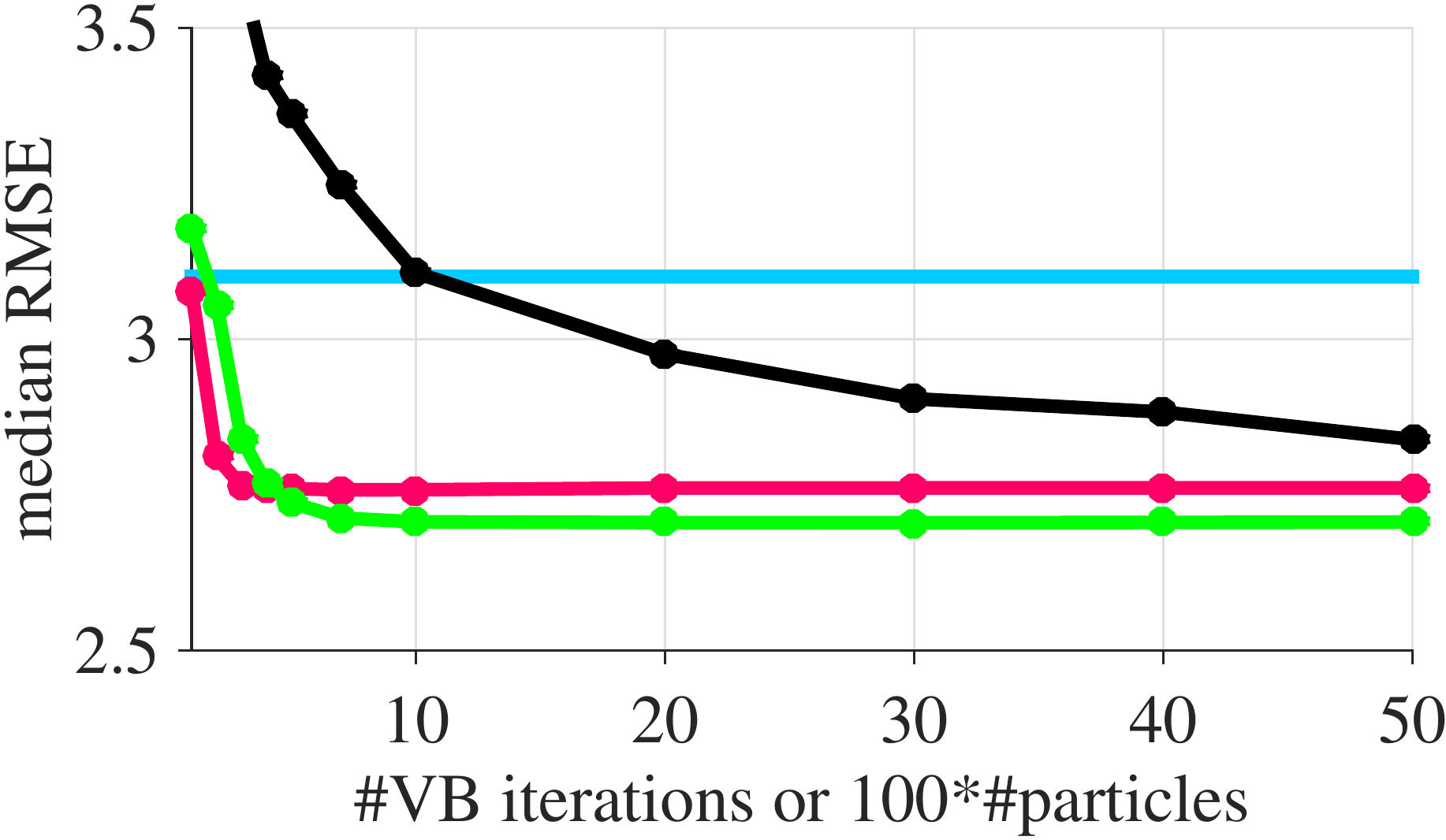}}
\hfil
\subfloat[$\delta=5$]{\includegraphics[width=0.48\columnwidth]{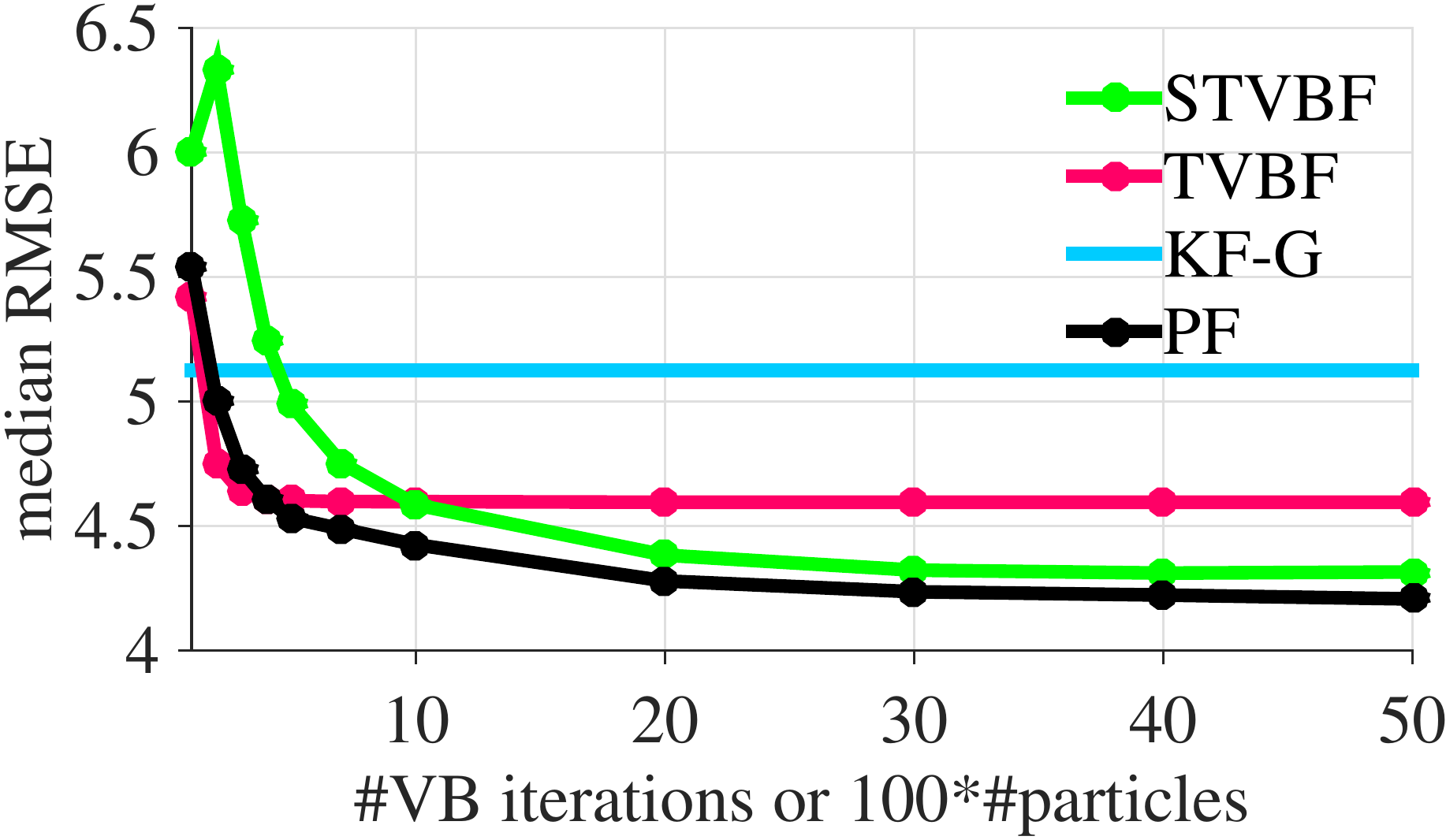}}
\vspace{-0.5mm}
\caption{Convergence of the STVBF with $q=10$. Ten STVBF iterations is enough to outperform TVBF. One additional VB iteration gives the same accuracy gain as 100 additional PF particles.
} \label{fig:time_vs_acc}
\end{figure}

Fig.\ \ref{fig:time_vs_acc} studies the convergence of the STVBF's VB iteration with $q=10$. The speed of convergence depends on the parameters of the model; the larger $\delta$, the slower convergence, and large $q$ and a high number of sensors can also increase the required number of iterations. The RMSE reduction is fastest for the first iterations, 10 iterations is enough to outperform TVBF, and after 30 iterations the RMSE reduction is negligible. Thus, the STVBF is slower than the TVBF that requires 5 iterations. In this example, one additional VB iteration gives the same accuracy gain as 100 additional PF particles. In the remaining numerical examples, STVBF's VB iteration is terminated after 30 iterations, and TVBF's after 10 iterations.

\begin{figure}[t]
\centering
\subfloat[$q=0.1$]{\includegraphics[clip,width=0.48\columnwidth]{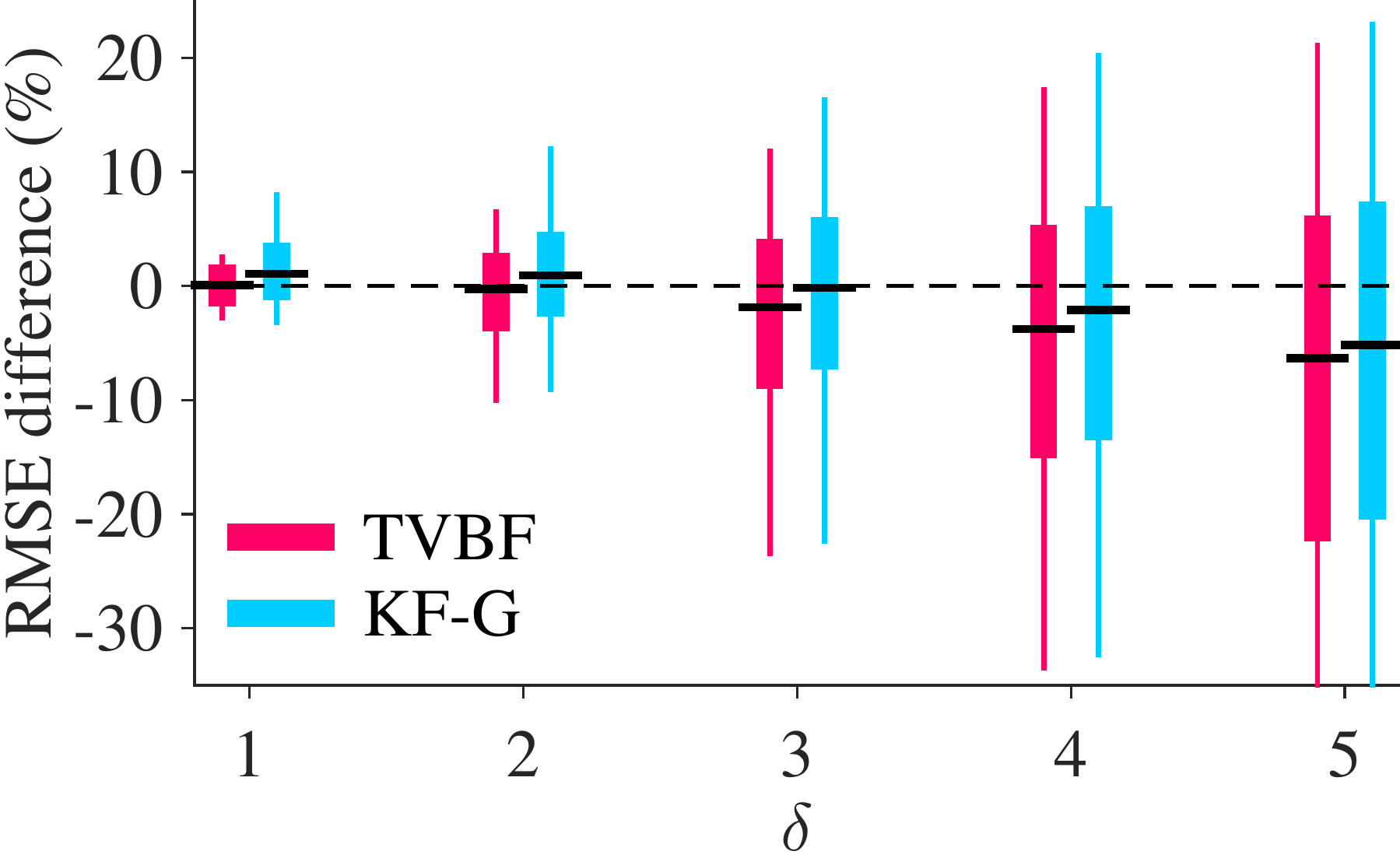}}
\hfil
\subfloat[$q=1$]{\includegraphics[width=0.48\columnwidth]{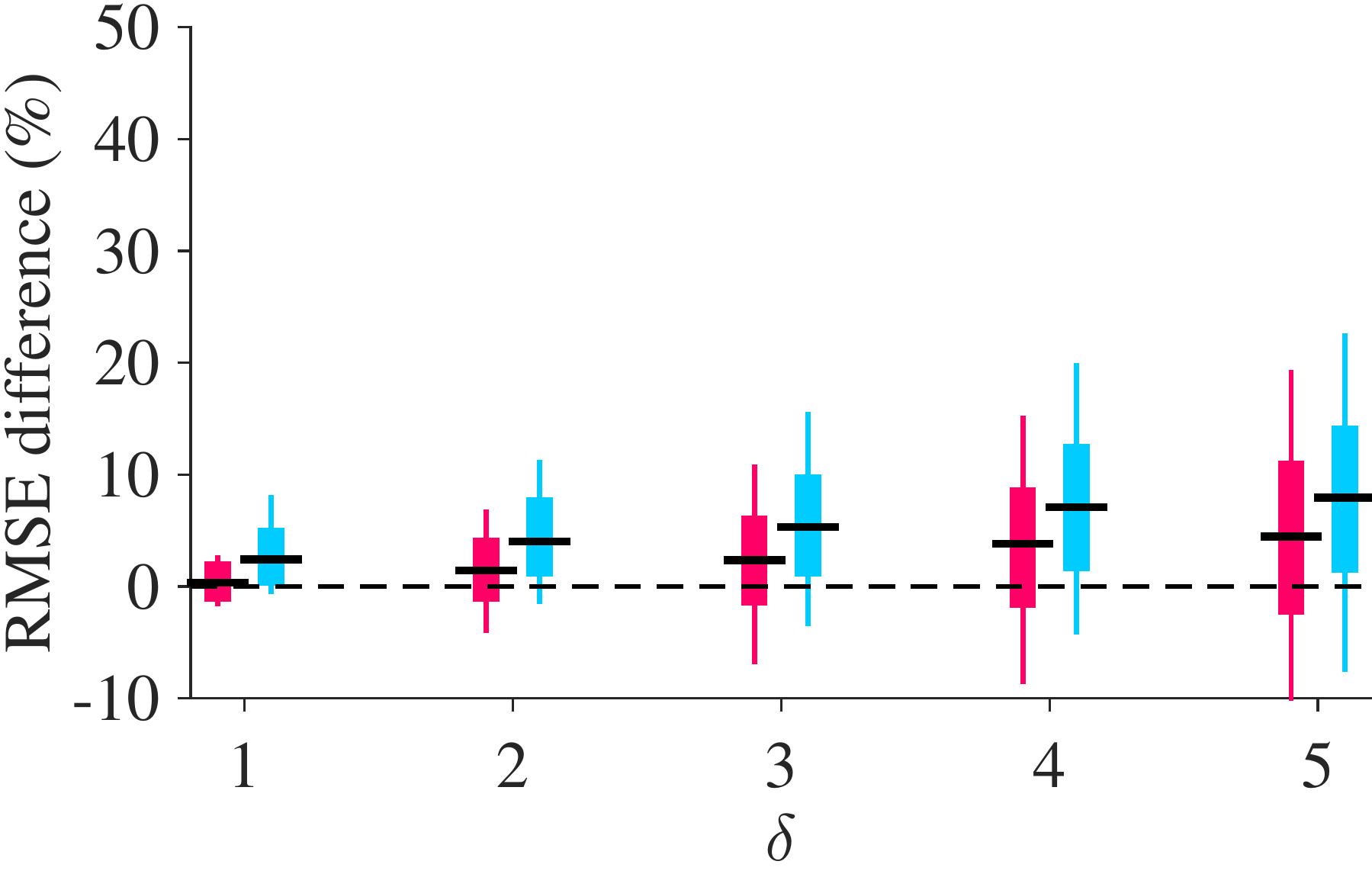}}
\\
\subfloat[$q=10$]{\includegraphics[width=0.48\columnwidth]{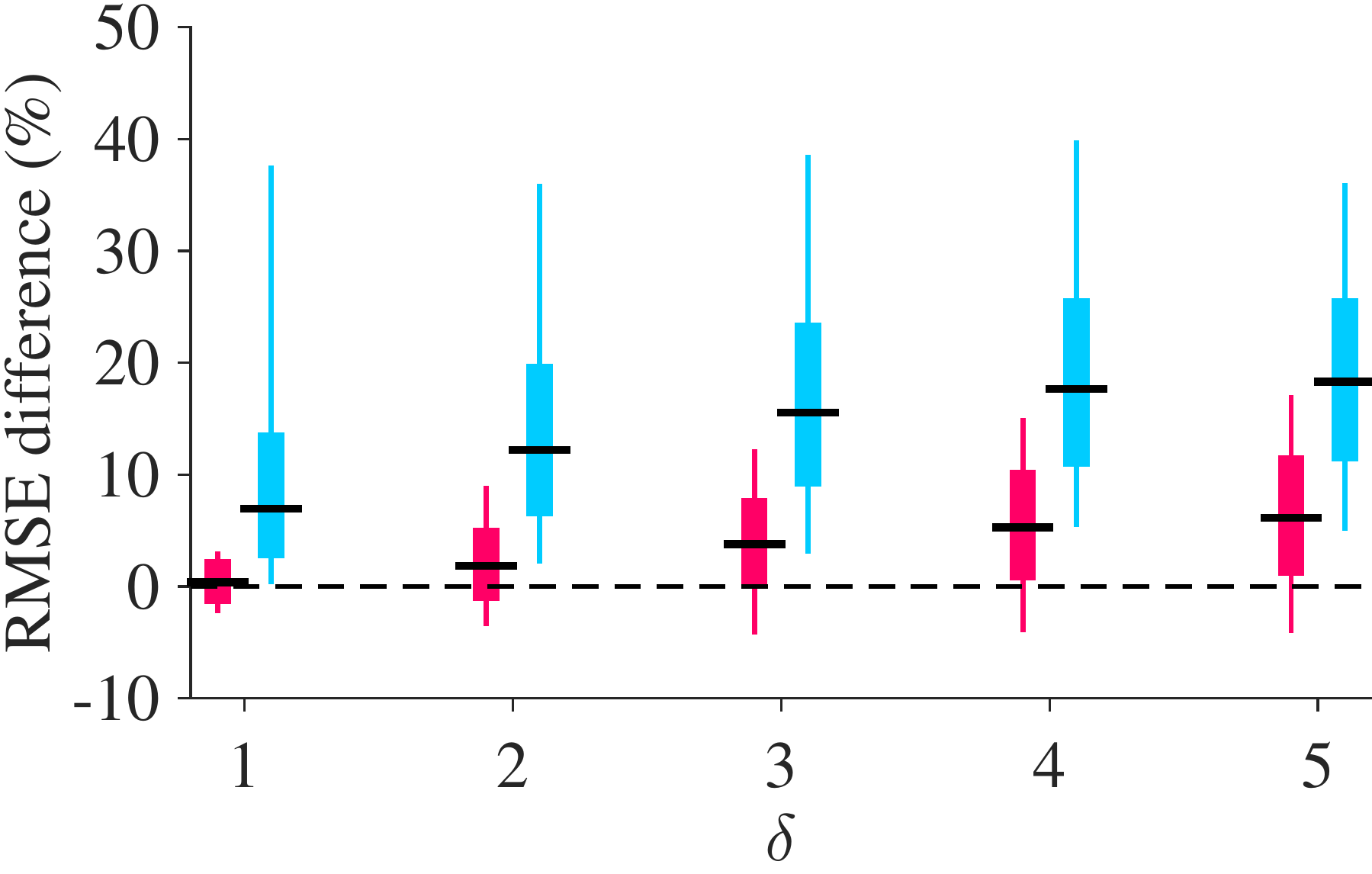}}
\hfil
\subfloat[$q=40$]{\includegraphics[width=0.48\columnwidth]{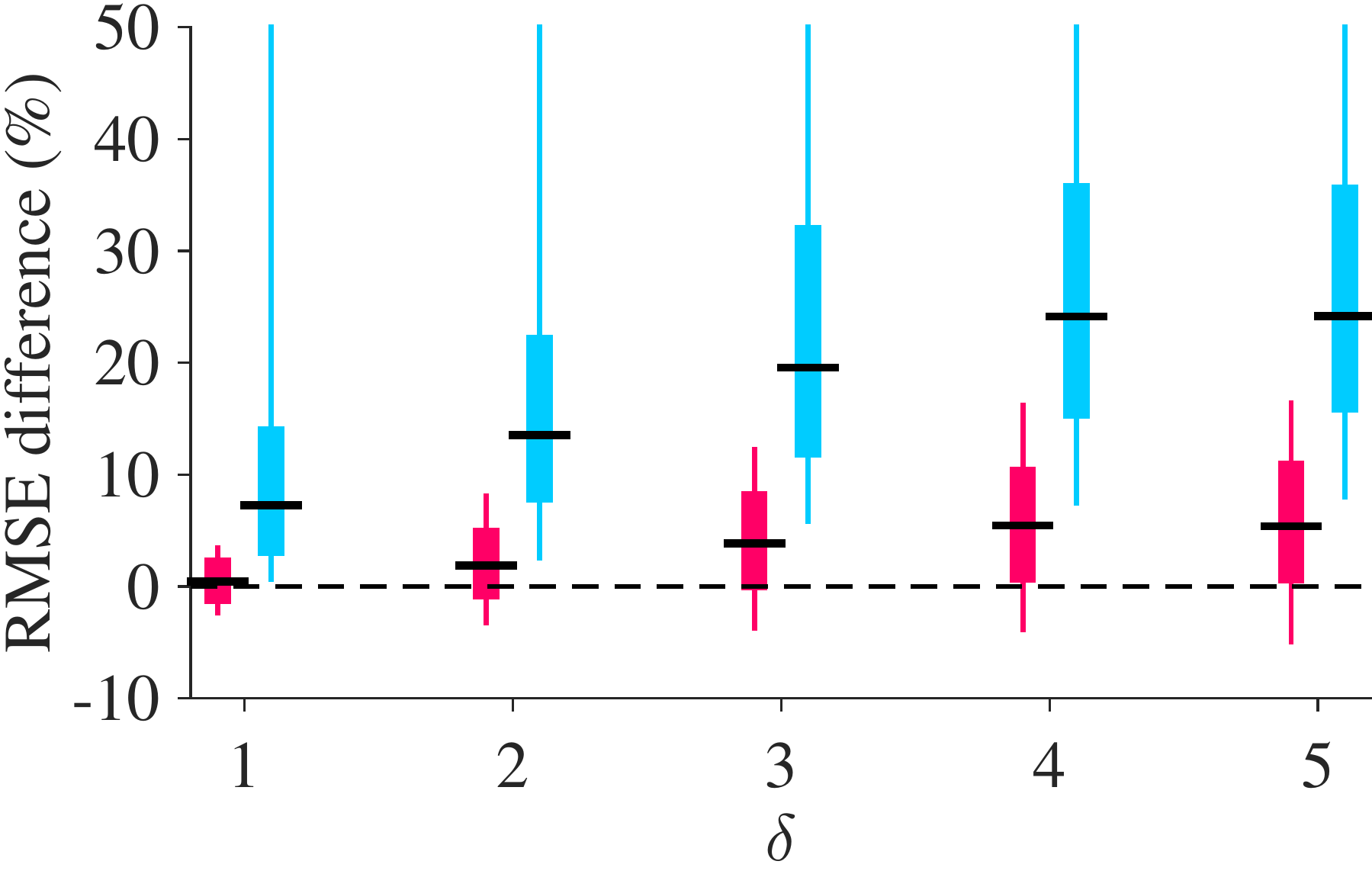}}
\vspace{-0.5mm}
\caption{RMSE differences per cent of the STVBF's RMSE. The proposed STVBF outperforms the comparison methods with skewed measurements when the signal-to-noise ratio is high enough.} \label{fig:rmse_path}
\end{figure}

Fig.\ \ref{fig:rmse_path} shows the distributions of the RMSE differences of the comparison methods from the STVBF's RMSE as percentages of the STVBF's RMSE. The levels of the boxes are 5\,\%, 25\,\%, 50\,\%, 75\,\%, and 95\,\% quantiles. With $q\geq1$, the STVBF outperforms the comparison methods in significant majority of the replications. 
The problems with $q=0.1$ are explained by the model structure: only sums of $x_k$ and $u_k$ are measured, so $x_k$ and $u_k$ are correlated \textit{a posteriori}, which makes the VB approximation underestimate the posterior variance \cite[Ch. 10.1.2]{Bishop2007}. The STVBF works well only when the process noise has enough dispersion to dominate in the prior's variance, i.e. when the signal-to-noise ratio (SNR) is not very low.

\subsubsection{Real-world noise}	
\begin{figure}[t]
\centering
\includegraphics[width=0.7\columnwidth]{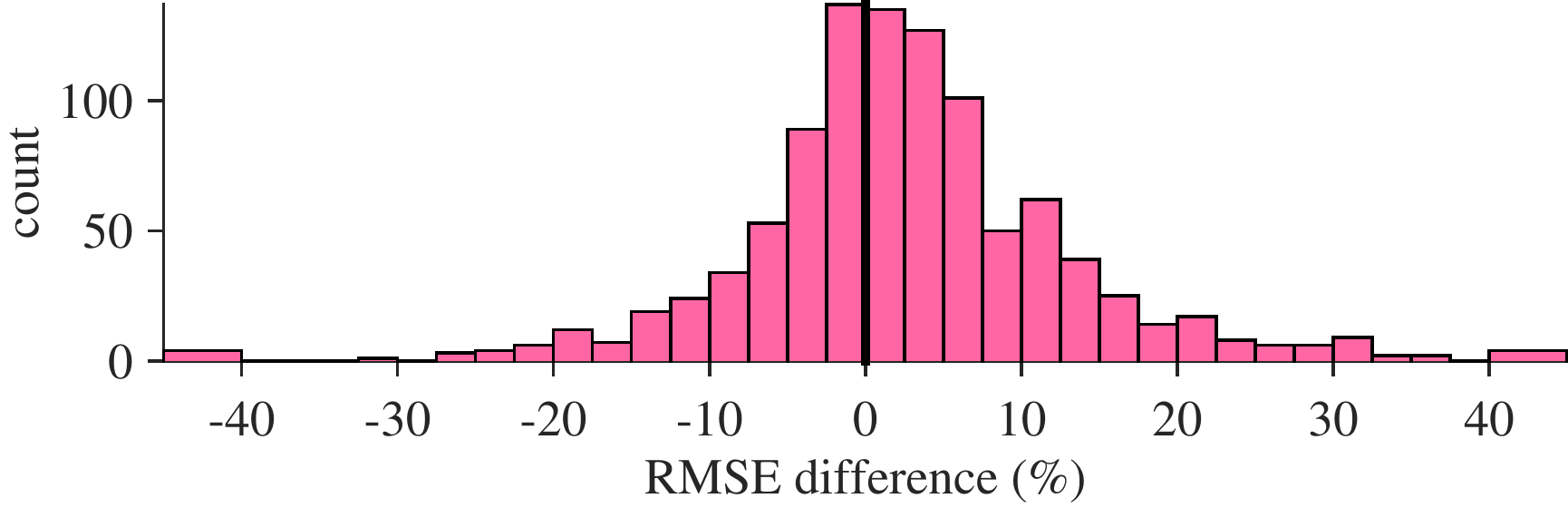}
\vspace{-2mm}
\caption{RMSE difference of TVBF per cent of the STVBF's RMSE with noise generated from real time-of-flight measurements' error histogram. STVBF has lower RMSE than the TVBF in 61\,\% of the 1000 replications.} \label{fig:rmsehist_uwbdata}
\end{figure}

The robustness of the STVBF is evaluated by generating the noise in Eq.\ \eqref{eq:measmodel} from the histogram distribution of the time-of-flight data set of Fig.\ \ref{fig:uwb_example} and using $q=10$. The histogram of the RMSE differences of TVBF from the RMSE of STVBF is in Fig.\ \ref{fig:rmsehist_uwbdata}. The proposed method has lower RMSE than the TVBF in 61\,\% of the 1000 Monte Carlo replications. This indicates that the proposed filter is robust to small deviations from the model that appear in real data.

\subsubsection{Evaluation of the smoother}	
\begin{figure}[h]
\centering
\subfloat[$q=1$]{\includegraphics[width=0.48\columnwidth]{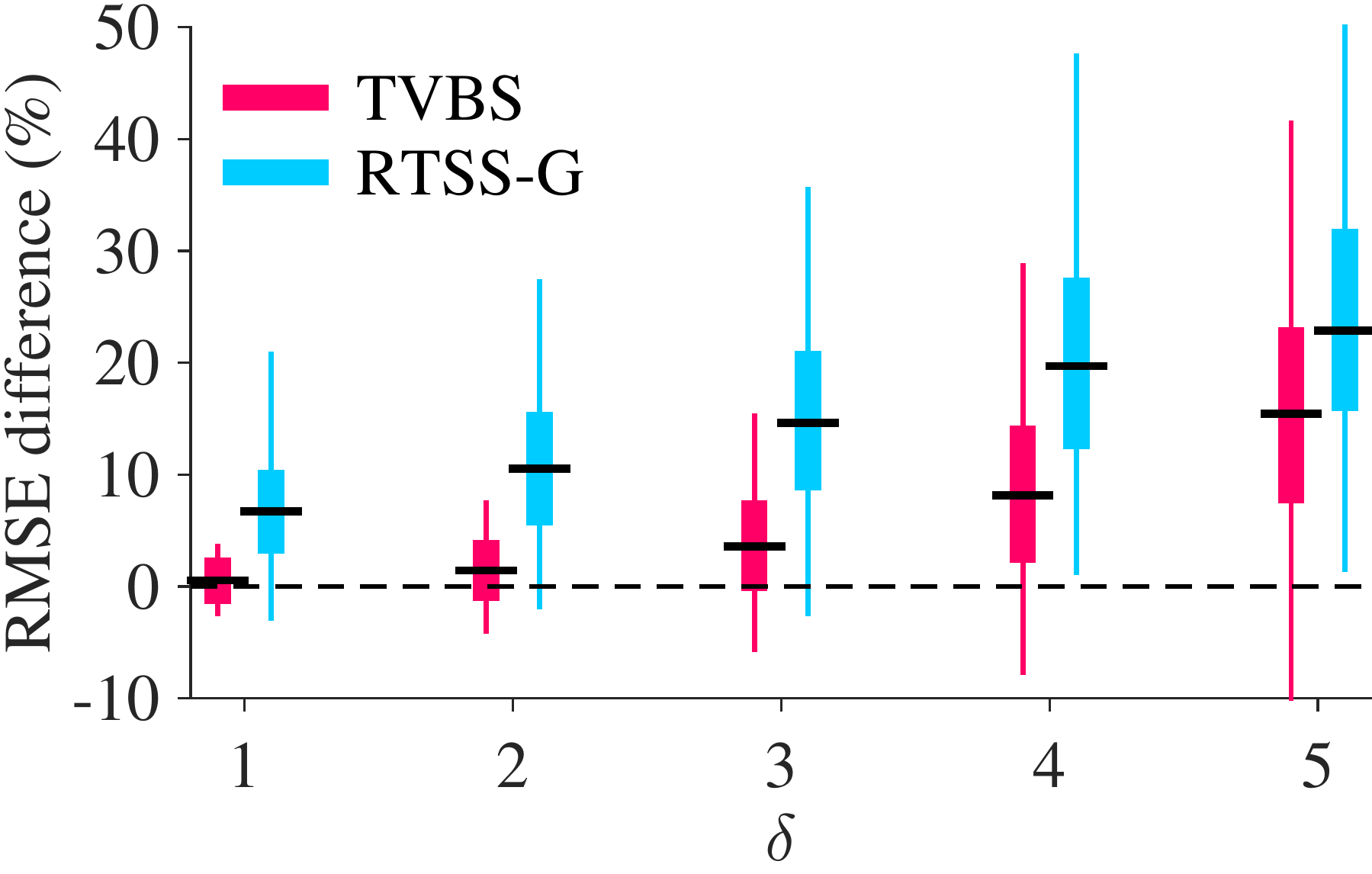}}
\hfil
\subfloat[$q=40$]{\includegraphics[width=0.48\columnwidth]{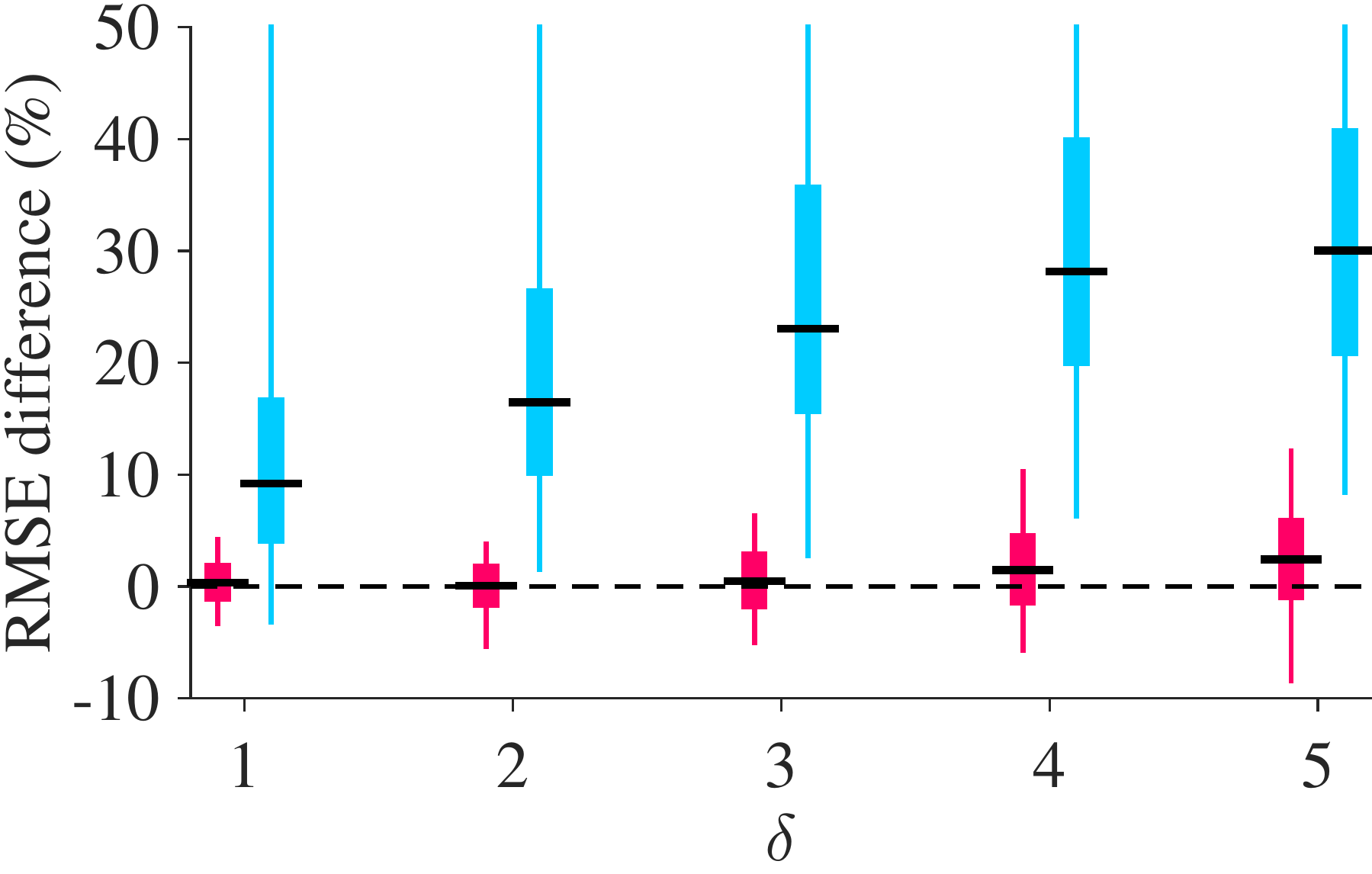}}
\vspace{-0.5mm}
\caption{Smoothers' RMSE differences per cent of the STVBS's RMSE. STVBS performs well also at low SNR, but difference to TVBS is smaller than the difference between the corresponding filters.} \label{fig:rmse_smoothing}
\end{figure}

The smoother versions of the compared algorithms are evaluated in the same simulation of Eq.\ \eqref{eq:measmodel} with skew-$t$ noise. The STVBS uses 30 and the TVBS 10 VB iterations, which were observed to provide convergence. Fig.\ \ref{fig:rmse_smoothing} shows that the STVBS outperforms the TVBS also at low SNR, but the percentile differences at high SNR are smaller than those of the corresponding filters.

%% file: conclusions.tex

\section{Conclusions} \label{sec:conclusions}

A filter and a smoother that take into account the skewness and heavy-tailedness of the measurement noise are proposed. The algorithms use the variational Bayes approximation. In the presented computer simulations the proposed methods outperform the conventional symmetric Kalman-type algorithms when skewness is present. The computational burden depends on the measurement dimension and model parameters. In the presented simulations the proposed filter has roughly 5 to 10 times the Kalman filter's computational cost.